\documentclass[twocolumn,showpacs,preprintnumbers,amsmath,amssymb]{revtex4}

\usepackage{graphicx}
\usepackage{dcolumn}
\usepackage{bm}
\usepackage{amssymb}
\usepackage{graphicx}
\usepackage{amsmath}
\usepackage{xspace}
\usepackage{natbib}

\begin{document}


\title{Distortion-Dependent Raman Spectra and Mode Mixing in $R$MnO$_3$ Perovskites}
\author{M. N. Iliev$^1$}
\author{M. V. Abrashev$^2$}
\author{J. Laverdi\`ere$^3$}
\author{S. Jandl$^3$}
\author{M. M. Gospodinov$^4$}
\author{Y.-Q. Wang$^1$}
\author{Y.-Y. Sun$^1$}
\affiliation{$^1$Texas Center for Superconductivity and Department
of Physics, University of Houston, Texas 77204-5002, USA}
\affiliation{$^2$Faculty of Physics, University of Sofia, 1164
Sofia, Bulgaria} \affiliation{$^3$D\'epartement de Physique,
Universit\'e de Sherbrooke, Sherbrooke, Canada J1K 2R1}
\affiliation{$^4$Institute of Solid State Physics, Bulgarian
Academy of Sciences, 1184 Sofia, Bulgaria}

\date{\today}

\begin{abstract}
The polarized Raman spectra of orthorhombic $R$MnO$_3$ series
($R$=La,Pr,Nd,Sm,Eu,Gd,Tb,Dy,Ho,Y) were studied at room
temperature. The variation of phonon frequencies with $R$ ionic
radius $r_R$ as a whole confirms the commonly accepted Raman line
assignments with two noticeable exceptions: (1) with decreasing
$r_R$ the stretching $A_g(1)$ and bending $A_g(3)$ modes strongly
mix for $R$=Sm to Tb, while for further decrease or $r_R$
($R$=Dy,Ho,Y) the $A_g(3)$ mode is observed at higher frequency
than $A_g(1)$ mode; (2) similar distortion-dependent mode mixing
takes place for the rotational $A_g(2)$ and O1($x$) [$A_g(7)$]
modes. The mode mixing is particularly strong for the $R$MnO$_3$
compounds with $r_R$ values close to the transition from A-type to
incommensurate sinusoidal antiferromagnetic ordering at low
temperatures. The frequency of rotational $A_g(2)$ and $A_g(4)$
modes scales to the angles of MnO$_6$ [101] and [010] rotations,
respectively, and could be used as a measure of their value.
\end{abstract}


\maketitle
\section{Introduction}
Until recently the $R$MnO$_3$ perovskites ($R$=rare earth,Y,Sc)
have been object of research mainly as parent materials of mixed
valence manganites exhibiting colossal magnetoresistivity
(CMR).\cite{Kusters89,Helmolt93,Jin94,Tokura94} In the past few
years, however, there is an increased interest in the complex
relationships among the lattice distortions, magnetism, dielectric
and transport properties of undoped $R$MnO$_3$.
\cite{Munoz01,Munoz02,Kimura03a,Kimura03,Goto04,Dabrowski05} In
particular, it has been found that with decreasing radius ($r_R$)
of $R$ ($R$=La to Eu) the transition temperature T$_N$ to A-type
antiferromagnetic (A-AFM) structure also decreases. With further
decrease of $r_R$ ($R$=Gd to Ho) the magnetic structure below
T$_N$ changes from A-AFM to incommensurate antiferromagnetic
one(IC-AFM) with sine-wave ordering of the Mn moments with
temperature-dependent wave vector $\vec{k_s}=(k_s,0,0)$ along the
$a$-axis (in $Pnma$ notations). At T$_{lock}$$<$T$_N$ an
incommensurate-to-commensurate (IC-CM) transition takes place and
$k_s$ locks at a value $< 0.4$.\cite{Kimura03} A transition to
commensurate E-type antiferromagnetic structure (E-AFM, $k_s =
0.5$) has been observed below T$_{lock}=26K$ only for HoMnO$_3$.
\cite{Munoz02} For TbMnO$_3$ and DyMnO$_3$ it was found that the
IC-CM transition is accompanied by a ferroelectric transition,
associated with lattice modulation in the CM
phase.\cite{Kimura03a,Goto04}. Large magnetodielectric effects
have also been reported for orthorhombic  HoMnO$_3$, and
YMnO$_3$.\cite{Lorenz04}

 Although
the role of lattice distortions in the interplay of magnetic,
dielectric and transport properties of $R$MnO$_3$ perovskites is
widely recognized, there are relatively few studies on the
variations of these distortions with
$R$.\cite{Alonso00,Dabrowski05} Raman spectroscopy can provide
significant additional information on the distortions and their
variations with both $R$ and temperature. The Raman active modes
in $R$MnO$_3$ ($7A_g+5B_{1g}+7B_{2g}+5B_{3g}$) are activated
exclusively due to the deviations from ideal perovskite structure
and in principle the activation of each Raman line can with definite certainty be
assigned to the value of one or two types of basic distortion (rotations of
MnO$_6$ octahedra around [101] or [010] directions, Jahn-Teller
distortion, or shift of $R$ atoms).\cite{Abrashev02} The first
assignment of the Raman lines in the spectra of $R$MnO$_3$ to
definite atomic vibrations has
been done by comparison of experimentally obtained Raman phonon
frequencies of LaMnO$_3$ and YMnO$_3$ with those predicted by
lattice dynamical calculations (LDC). \cite{Iliev98} Although for
most of the experimentally observed Raman lines the frequencies
have been in good agreement with those predicted by LDC, some
uncertainties about the phonon line assignment remained. Indeed,
some Raman lines of same symmetry have close frequencies and due
to the approximations of the LDC model one reasonably expects some
difference between measured and calculated values. The assignment
of the two $A_g$ Raman lines between 450 and 520~cm$^{-1}$, where
are the $A_g$ modes corresponding to anti-stretching(AS) and
bendings(B) of MnO$_6$ octahedra, has been challenged by
Martin-Carron et al.\cite{Martin-Carron02} in a Raman study of
several $R$MnO$_3$ perovskites ($R$=La,Pr,Nd,Tb,Ho,Er,Y). Based on
the fact that the Mn-O distances exhibit only small changes within
the $R$MnO$_3$ series, they assigned the AS band observed near
480-490~cm$^{-1}$ to the anti-stretching mode, while the B band
which shifts from $\approx 450$~cm$^{-1}$ in LaMnO$_3$ to $\approx
530$~cm$^{-1}$ in ErMnO$_3$ has been assigned to bending mode(s).
An obvious problem of this re-assignment is the "crossing" of the
AS and B modes, which are of the same A$_g$ symmetry.

In this work we present results of a detailed study of polarized
Raman spectra of $R$MnO$_3$ ($R$=La,Pr,Nd,Sm,Eu,Gd,Tb,Dy,Ho,Y) at
room temperature. The comparison of variations with $R$ of phonon
frequencies and relative intensities of most pronounced $A_g$ and
$B_{2g}$ modes shows that some corrections of previous phonon mode
assignment are needed. The strong mixing and spectral weight
transfer between $A_g$ anti-stretching (AS) and bending (B) modes
in the 500~cm$^{-1}$ region, most clearly observed for $R$=Eu,Gd
and Tb, correlates with changes of magnetic structure. Similar
distortion-dependent mode mixing is observed for the rotational
$A_g(2)$ and O1($x$) [$A_g(7)$] modes.

\section{Samples and Experimental}
PrMnO$_3$, NdMnO$_3$ and SmMnO$_3$ samples were grown by the
floating zone method described in Ref.\cite{Balbashov96}. Single
crystals of DyMnO$_3$ of average sizes $1 \times 1 \times
2$~mm$^3$ were prepared following a procedure to be described
elsewhere \cite{Gospodinov05}. Orthorhombic YMnO$_3$, HoMnO$_3$
samples were prepared under pressure as described in
Ref.\cite{Lorenz04}. For $R$=Eu,Gd,Tb the precursor powders were
heated at 1130-1160$^\circ C$ in O$_2$ for 24 hours.  Although the
samples (except for DyMnO$_3$) were polycrystalline, the size of
the constituting microcrystals was larger than the laser spot. It
was possible to select microcrystals with crystallographic
orientation such that the spectra in parallel and crossed
scattering configurations were dominated by $A_g$ or $B_{2g}$
contributions, respectively. For exact separation of the $A_g$ and
$B_{2g}$ spectra the SPECTRA SUBTRACT program of the GRAMS AI
software package was used.

The Raman spectra were measured at room temperature in
backscattering configuration using two different spectrometers:
Labram-800 and Jobin Yvon HR640, both equipped with microscope and
liquid-nitrogen-cooled CCD detector. The spectra obtained with
514.5~nm (Ar$^+$) and 632.8~nm (He-Ne) were practically identical.

\section{Results and Discussion}
The $A_g$ and $B_{2g}$ spectra of $R$MnO$_3$ as obtained at room
temperature are shown in Figure 1. With same scattering
configuration the absolute value of Raman intensities were
increasing with decreasing R ionic radius ($r_R$). For better
comparison the $A_g$ and $B_{2g}$ spectra were normalized to the
integrated intensities of $A_g(1)+A_g(3)$ and $B_{2g}(1)$ lines,
respectively. At least five $A_g$ and four $B_{2g}$ lines are
clearly pronounced and the variations of their parameters with $R$
can be followed with a good accuracy. The phonon line frequencies
are given in Table~I.

\begin{figure}[htbp]
\includegraphics[width=7cm]{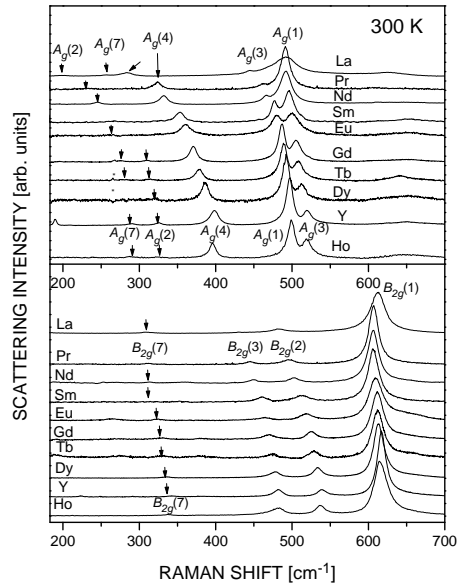}
\caption{$A_g$ and $B_{2g}$ Raman spectra of $R$MnO$_3$ at room
temperature. The phonon mode notations correspond to those of
Refs.\cite{Abrashev02,Iliev98}. The $A_g$ and $B_{2g}$ spectra are
normalized to the integrated intensity of $A_g(1)+A_g(3)$ and
$B_{2g}(1)$ lines, respectively.}
\end{figure}

\begin{table*}[htb]
\caption{Frequencies (in cm$^{-1}$) of the experimentally observed
$A_g$ and $B_{2g}$ Raman lines of $R$MnO$_3$ at room temperature.}
\begin{tabular}{ccccccccccccc}
\hline
Mode &Main atomic motions &La&Pr &Nd &Sm &Eu &Gd &Tb &Dy &Ho &Y&Basic distortion  \\
\hline
$A_g(1)\rightarrow A_g(3)$&O2 anti-stretchng $\rightarrow$ MnO$_6$ bending&490.7 &491.2 &491.7 &496.1 &501.0 &506.0 &509.0 &513.4 &520.0 &520.6&JT,[010]$\rightarrow$[101] \\
 $A_g(3)\rightarrow A_g(1)$&MnO$_6$ bending $\rightarrow$ O2 anti-stretching&444.8 &461.8 &465.6 &476.7 &479.0 &486.4 &488.8 &492.1 &498.9 &496.7&[101]$\rightarrow$JT,[010] \\
 $A_g(4)$&out-of-phase MnO$_6$ $x$-rotations &283.6 &323.7 &331.8 &352.9 &360.8 &370.5 &378.2 &386.0 &395.9 &398.1& [101]\\
 $A_g(2)\rightarrow A_g(7)$&in-phase MnO$_6$ $y$-rotations$\rightarrow$ O1($x$) &256.8 & & & & &309.8 &314.6 &319.5 &323.8 & 325.3&[010]$\rightarrow R$-shift \\
  $A_g(7)\rightarrow A_g(2)$&O1($x$) $\rightarrow$in-phase MnO$_6$ $y$-rotations &198.1 &231.8 &245.0 &266.1 & &276.1 &280.5 & &287.8 & 289.4 &$R$-shift $\rightarrow$[010]\\
 \hline \\

 $B_{2g}(1)$&in-plane O2 stretching &612.3 &606.7 &606.2 &606.6 &609.5 &611.7 &612.2 &613.5 &615.0 &617.3& JT\\
$B_{2g}(2)$&in-phase O2 "scissors'-like &482.5 &495.7 &502.6 &513.1 &518.4 &525.0 &528.4 &533.6 &537.0 &539.2&[010] \\
 $B_{2g}(3)$&out-of-phase MnO$_6$ bending &430.0 &444.8 &449.1 &460.9 &464.6 &468.9 &473.7 &477.5 &481.1 &481.5&[101] \\
 $B_{2g}(7)$&O1($z$) &309.7 &311.5 &313.2 & &323.8 &329.2 &330.9 &335.6
 & &342.0 & $R$-shift

\end{tabular}
\end{table*}

\begin{figure}[htbp]
\includegraphics[width=7cm]{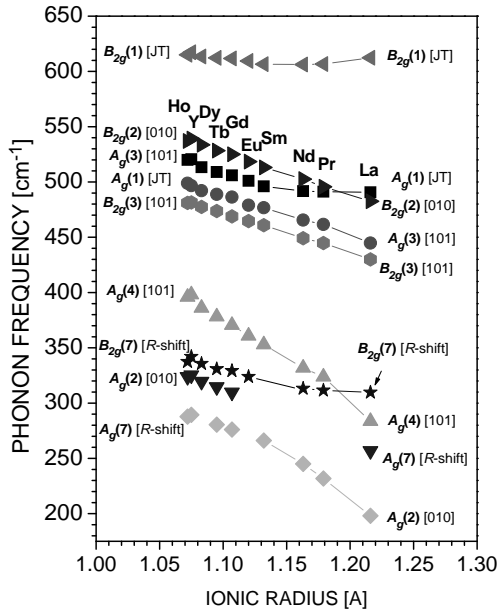}
\caption{Variations of phonon mode frequencies with $R$ ionic
radius $r_R$.}
\end{figure}

The correlation between structural and physical parameters of
$R$MnO$_3$ perovskites may be described as a function of different
parameters, such as the ionic radius $r_R$, tolerance factor $t$
or the averaged tilt angle $\Psi$. The relation between these
parameters, however, is close to linear\cite{Dabrowski05} and any
one of them can be used for characterization of phonon mode
frequency variations. Figure 2 shows the dependence of Raman
phonon frequencies on $r_R$. Most Raman frequencies $\omega_i$
exhibit nearly linear increase with $r_R$, but the slope $\Delta
\omega_i / \Delta r_R$ depends mainly on the type of the mode
(stretching,bending or rotational) as well on the type of
distortion that has activated the corresponding mode in the Raman
spectrum (MnO$_6$ rotations around $[010]_c$ or $[101]_c$ cubic
axes, Janh-Teller(JT) distortion, and $R$-shift).\cite{Abrashev02}

\subsection{Basic distortions and Raman modes in $R$MnO$_3$}
In the ideal cubic ABO$_3$ perovskite (space group $Pm\bar{3}m$,
$Z=1$) all atoms are at centrosymmetrical sites with fixed
coordinates and there are no Raman allowed modes. The structure of
stoichiometric $R$MnO$_3$, described at room temperature by the
$Pnma$ space group ($Z=4$), can be considered as orthorhombically
distorted superstructure of the ideal perovskite with $[100]_o$,
$[010]_o$, and $[001]_o$ directions coinciding with the $[101]_c$,
$[010]_c$ and $[\bar{1}01]_c$ directions of the parent cubic
structure. In the $Pnma$ structure the atoms occupy four
non-equivalent atomic sites ($R$,Mn,O1,O2), of them only the Mn
site is a center of symmetry. From symmetry considerations only 5
of the 12 atomic coordinates are fixed [$R$-1,Mn-3,O1-1], while
the remaining 7 variable coordinates can be considered as the
lattice degrees of freedom producing the above mentioned lattice
distortion. These distortions give rise to 24
($7A_g+5B_{1g}+7B_{2g}+5B_{3g}$) Raman-allowed phonon modes, which
in turn can be assigned to their distortions of origin ([010]-,
[101]-rotation, JT, or $R$-shift). With decreasing $r_R$ the
distortions increase and most of the bond lengths shorten, which
reasonably results in twofold effect: enhanced Raman intensity and
hardening of Raman frequencies.

At first sight the [010] and [101] rotations can directly be
obtained from the O2-Mn-O2 and O1-Mn-O1 angles, which practically
have equal values.\cite{Alonso00,Dabrowski05} As shown by Abrashev
et al.\cite{Abrashev02}, however, even in the case of equal O-Mn-O
angles, the [010] and [101] rotation angles will be different.
They are related to the deviations ($x_i,y_i,z_i$, $i$-atomic
index) of $Pnma$ atomic coordinates from those in a hypothetical
undistorted perovskite structure as
$$\Psi_{[010]} = \arctan(2|x_{O2}-z_{O2}|)$$
$$\Psi_{[101]} = \arctan(2^{5/2}|y_{O2}|) \approx \arctan(2^{3/2}|z_{O1}|) $$

The so defined atomic deviations can also be used to describe the
other two basic distortions, namely, the JT distortion,
characterized by the relative difference $D_{JT}= | \Delta
d_{Mn-O2}| /<d_{Mn-O2}>$ of the two pairs of Mn-O2 distances, and
the shift $D_R(x)$ of the $R$-atoms in $x$-direction from their
positions in an undistorted perovskite. In the case of small
distortions:
$$D_{JT} = 2 |x_{O2} + z_{O2} | $$
$$ D_{R(x)} = 2x_R$$

The variations of the basic distortions, calculated using the
available data for atomic positions in
$R$MnO$_3$\cite{Alonso00,Dabrowski05} are given in Table~II.

\begin{table*}[htb]
\caption{Variations with $R$ of the four basic distortions of the
$Pnma$ structure. The data for SmMnO$_3$ and GdMnO$_3$ have been
obtained by the fit of corresponding experimental data for rest
$R$MnO$_3$ compounds.\cite{Dabrowski05,Alonso00}}
\begin{tabular}{cccccccccccc}
\hline
Distortion &Measure &La&Pr &Nd &Sm &Eu &Gd &Tb &Dy &Ho &Y  \\
\hline

[010]& $\Psi_{[010]}$ [deg] & 9.23 &11.43  &12.11  &13.07  & 13.35  & 13.60  &13.81  & 14.00  &14.16  &14.11  \\

[101]& $\Psi_{[101]}$ [deg] &  12.17&13.53  &14.08  & 15.10  &15.49  &15.86  &16.24  &16.59  &16.90  &16.82  \\

JT & $| \Delta d_{Mn-O2}|/<d_{Mn-O2}>$&0.0647  &0.0665  &0.0652  & 0.0612  &0.0597  &0.0582  &0.0574  & 0.0571  & 0.0575 &0.0573 \\

$R$-shift& $2x_R$ &0.097  &0.129  &0.138  & 0.151 &0.155  &0.159  &0.166  &0.167  &0.169  &0.168  \\
\hline
$<$Mn-O2$>$ Refs.\cite{Dabrowski05}  & \AA &2.044  &2.063  &2.067  &  &2.066  &  &  &2.065  &  &  \\
$<$Mn-O2$>$ Refs.\cite{Alonso00}     & \AA &2.043  &2.059  &2.062  &  &  &  &2.063  &2.063  &2.064  &2.052  \\
\hline
\end{tabular}
\end{table*}
\subsection{Pure and mixed stretching (Jahn-Teller) modes}

The frequency of a mode involving mainly stretching vibrations of
O2 atoms in the $xz$ planes is determined by the Mn-O2 distances.
The short and long Mn-O2 bond lengths vary with $r_R$ within 0.5\%
and 1.1\%, respectively\cite{Alonso00,Dabrowski05} and therefore
one expects that $\omega_{stretch} \propto d^{3/2}_{Mn-O2}$ will
change by no more than 1.5\%. As illustrated in Figure 3, this is
exactly the case for the $B_{2g}(1)$ mode at 606-617~cm$^{-1}$,
which has been assigned to the in-phase O2
stretching.\cite{Iliev98}. The minimum of $B_{2g}(1)$ frequency
near R=Nd,Sm reflects the fact that for these $R$'s the averaged
Mn-O distance has a maximum.

Based on the same bond-length considerations, one could expect
that the Raman line of $A_g$ symmetry at 490-495~cm$^{-1}$,
corresponding to the in-phase anti-stretching vibrations of O2 in
the $xz$ plane,\cite{Iliev98} would behave in a similar way and
its frequency would remain nearly constant through the whole
$R$MnO$_3$ series. As seen from Fig.4, this is obviously not the
case. The data of Figs.~1 and 4 clearly show a classic example for
mixing of modes of same symmetries and close frequencies. For
large $r_R$ ($R$=La,Pr,Nd) the lines at 490-492 and
445-465~cm$^{-1}$ can be considered as pure stretching($A_g(1)$)
and bending($A_g(3)$) modes, respectively. With decreasing $r_R$
the $A_g(3)$ frequency increases and approaches that of $A_g(1)$,
resulting in strong mode mixing evidenced by mode repulsion and
transfer of intensity. The mixing is most strongly pronounced for
$R$=Sm to Tb where the two modes are of comparable intensities and
involve both stretching and bending atomic motions. With further
decrease of $r_R$ ($R$=Dy,Ho,Y) the modes become less mixed but
now the higher mode is dominated by MnO$_6$ bending
[$A_g(3)$-type] motions and the lower one is $A_g(1)$-type.

Two phonon modes of same symmetries and close frequencies can be
considered as coupled quantum oscillators. In a good approximation
their frequencies are given by
$$\omega_{1,2} = \frac{\omega ' + \omega ''}{2} \pm \sqrt{\frac{(\omega ' - \omega '')^2}{4} +  \frac{V^2}{4} }$$
where $\omega '$ and $\omega ''$ are the mode frequencies without
coupling and $V$ is the coupling constant. It is reasonable to
assume that without coupling the $A_g(1)$ and $A_g(3)$ mode
frequencies would depend on $r_R$ in the same way as the
$B_{2g}(1)$ and $B_{3g}(3)$ modes, respectively, as both modes of
the $A_g(1)$-$B_{2g}(1)$ (stretching) and $A_g(3)$-$B_{2g}(3)$
(bending) pairs are activated by the same basic distortion and
have similar shape. Therefore, as shown with dash lines in Fig.4,
the $\omega '(r_R)$ dependence for the uncoupled $A_g(1)$ mode can
be approximated by scaling the experimental $B_{2g}$ values,
whereas that of uncoupled $A_g(3)$ mode [$\omega ''(r_R)$], like
in the case of $B_{2g}(3)$, can be approximated by a linear
dependence. At the crossing point $(r_R \approx 1.125$~\AA),
$\omega' (1.125) = \omega '' (1.125)$, which gives $V = \omega_1
(1.125) - \omega_2 (1.125) \approx 20$~cm$^{-1}$.

\begin{figure}[htbp]
\includegraphics[width=7cm]{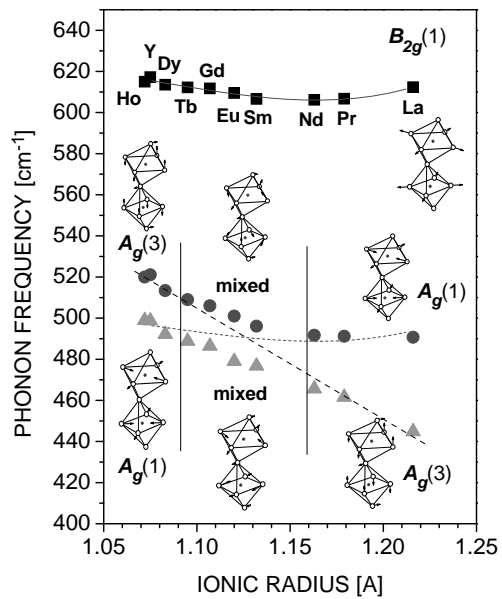}
\caption{Variations with $r_R$ of the frequencies of $B_{2g}(1)$,
$A_g(3)$ and $A_g(1)$ modes. The dashed curves show the expected
$\omega ' (r_R)$ and $\omega '' (r_R)$ dependences for pure
stretching and bending modes.}
\end{figure}
\subsection{Pure and mixed rotational ("soft") modes}
The frequency vs $\Psi_{[101]}$  dependence for the $A_g(4)$ mode
involving mainly MnO$_6$ $x$-rotations, shown in Fig.4 with full
circles, is close to proportionality. Expectedly, this mode
exhibits "soft mode" behavior with $\omega (\Psi ) \rightarrow 0$
for $\Psi \rightarrow 0$, the slope of
$\omega^{A_g(4)}(\Psi_{[101]})$  being 23.5~cm$^{-1}$/deg.
Calculations of lattice dynamics for the two end compounds
LaMnO$_3$ and YMnO$_3$ have shown that the two weak $A_g$ modes at
lower frequencies are strongly mixed and involve both in-phase
MnO$_6$ $y$-rotations [$A_g(2)$] and O1(x)-motions [$A_g(7)$]. The
higher of these modes is $A_g(7)$-like in LaMnO$_3$, but
$A_g(2)$-like in YMnO$_3$. Assuming that the higher mode remains
$A_g(2)$-like in other compounds with small $r_R$ (HoMnO$_3$,
DyMnO$_3$, TbMnO$_3$, GdMnO$_3$), one finds that the frequency vs
$\Psi_{[010]}$  dependence for the partly mixed $A_g(2)$ mode
(shown with empty squares) is practically the same as that for the
pure $A_g(4)$ mode. Interestingly, the frequencies of [111]-
rotational modes of structurally different rhombohedral LaMnO$_3$
and LaAlO$_3$ fit to these dependences, which is an indication for
a quite general relationship between the angle of rotation and
frequency of the rotational "soft" modes in perovskitelike
compounds, independent on the direction of the rotational axis.

\begin{figure}[htbp]
\includegraphics[width=7cm]{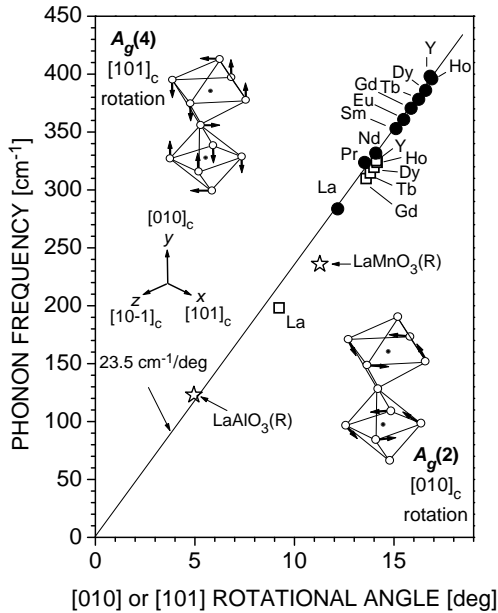}
\caption{Variations with the tilt angle $\Psi_{[101]}$ or
$\Psi_{[010]}$ of the frequencies of $A_g(4)$ and $A_g(2)$ modes,
involving mainly MnO$_6$ rotations.}
\end{figure}

\subsection{$R$-shift modes}
The weak $A_g(7)$ and $B_{2g}(7)$ involve mainly O1 motions in the
$xz$ plane, activated in the Raman spectrum due to the shift of
$R^{3+}$ ions from their positions in an ideal perovskite. It
follows from Fig.2, Table~I and Table~II that in the case of pure
$R$-shift mode [$B_{2g}(7)$], despite of strongly increasing
$R$-shift distortion between $R$=La and Y, the corresponding
increase of mode frequency is relatively modest (see Table~II). It
is plausible to expect similar dependence on $r_R$ for the
$A_g(7)$ mode, which explains the above discussed observation that
the frequency vs $r_R$ dependence of the mixed $A_g(2)+A_g(7)$
modes is governed mainly by their rotational [$A_g(2)$] component.

\section{Conclusion}
The comparative study of polarized Raman spectra of orthorhombic
$R$MnO$_3$ series ($R$=La,Pr,Nd,Sm,Eu,Gd,Tb,Dy,Ho,Y) shows that
the variations of lattice distortions with $r_R$ affect
significantly both the phonon frequencies and the shape of some
Raman phonon modes. The strong mixing of phonon modes involving
in-plane and out-of-plane oxygen motions may become a clue for
understanding the change with $r_R$ of magnetic ordering at low
temperatures. The established proportionality between the
frequency of a rotational mode and the angle of corresponding
rotational distortion may be used for material characterization.

\acknowledgments This work is supported in part by the State of
Texas through the Texas Center for Superconductivity and  the
National Science and Engineering Research Council of Canada. The
work of M.M.G. is supported by the Bulgarian National Research
Fund (Project F-1207).


\begin{thebibliography}{99}
\bibitem{Kusters89} R.M. Kusters, J. Singleton, D.A. Keen,
R. McGreevy, and W. Hayes, Physica B {\bf 155}, 362 (1989).
\bibitem{Helmolt93} R. von Helmolt, J. Wecker, B. Holzapfel, L. Schultz,
and K. Samwer, Phys.\ Rev.\ Lett.\ {\bf 71}, 2331 (1993).
\bibitem{Jin94}S. Jin, T. H. Tiefel, M. McCormack, R. A. Fastnacht,
R. Ramesh, and L. H. Chen, Science {\bf 64},413 (1994).
\bibitem{Tokura94} Y. Tokura, A. Urushibara, Y. Moritomo, T. Arima,
A. Asamitsu, G. Kido, and N. Furukawa, Science {\bf 63}, 3931
(1994).

\bibitem{Munoz01} A. Munoz, M. T. Cas\'ais, J. A. Alonso, M. J. Martinez-Lope, J. L.
Martinez, and M. T. Fern\'andez-D\'iaz, Inorg. Chem. {\bf 40},
1020 (2001).
\bibitem{Munoz02} A. Munoz, J. A. Alonso, M. T. Casais, M. J. Martinez-Lope,
J. L. Mart\'inez, and M. T. Fern\'andez-D\'iaz,  J. Phys.:
Condens. Matter {\bf 14}, 3285 (2002).
\bibitem{Kimura03a} T. Kimura, T. Goto, H. Shintani, K.~Ishizaka,
T. Arima, and Y.~Tokura, Nature {\bf 426}, 55 (2003).

\bibitem{Kimura03}T. Kimura, S. Ishihara, H. Shintani, T. Arima,
K. T. Takahashi, K. Ishizaka, and Y. Tokura, Phys.\ Rev.\ B {\bf
68}, 060403(R) (2003).
\bibitem{Goto04} T. Goto, T. Kimura, G. Lawes, A. P. Ramirez,
and Y. Tokura, Phys.\ Rev.\ Lett.\ {\bf 92}, 257201 (2004).

\bibitem{Dabrowski05} B. Dabrowski, S. Kolesnik, A. Baszczuk,
O.~Chmaissem, T.~Maxwell, and J.~Mais, J.\ Solid State Chem.\ {\bf
178}, 629 (2005).

\bibitem{Lorenz04} B. Lorenz, Y.~Q.~Wang, Y.~Y.~Sun, and
C.~W.~Chu, Phys.\ Rev.\ B {\bf 70}, 212412 (2004).

\bibitem {Alonso00} J. A. Alonso, M. J. Mart\'{i}nez-Lope,
M.~T.~Casais, and M.~T.~Fern\'{a}ndez-D\'{i}az, Inorg.\  Chem {\bf
39}, 917 (2000).
\bibitem{Abrashev02} M. V. Abrashev, J. B\"{a}ckstr\"{o}m,
L.~B\"{o}rjesson, V.~N.~Popov, R.~A.~Chakalov, N.~Kolev,
R.-L.~Meng, and M.~N.~Iliev, Phys.\ Rev.\ B {\bf 65}, 184301
(2002).

\bibitem{Iliev98}  M. N. Iliev, M. V. Abrashev, H. G. Lee, V. N. Popov, Y. Y. Sun,
C. Thomsen, R. L. Meng, and C. W. Chu,  Phys. Rev. B {\bf 57},
2872 (1998).

\bibitem{Martin-Carron02} L. Mart\'in-Carr\'on, A. de Andr\'es, M. J. Martinez-Lope,
M. T. Casais, and J. A. Alonso,  Phys. Rev. B  {\bf 66}, 174303
(2002).

\bibitem{Balbashov96} A. M. Balbashov, S. G. Karabashev,
Y. A. M. Mukovskii, and S. A. Zverkov, J.\ Cryst.\ Growth {\bf
167}, 365 (1996).

\bibitem{Gospodinov05} M. M. Gospodinov et al, unpublished
results.



\end{thebibliography}
\end{document}